\documentclass[aps,twocolumn]{revtex4} 
\usepackage{amssymb,amsmath,amscd}
\usepackage{graphicx} 
\newcommand{\etal}{\mbox{{\em et al}.\ }}                                 

\begin{document}
\title[]{
Growing Perfect Decagonal Quasicrystals by Local Rules
}
\author{Hyeong-Chai Jeong}
%\email{hcj@sejong.ac.kr} 
\affiliation{
Department of Physics,
Sejong University, Seoul 143-747, Korea, \\
Asia Pacific Center for Theoretical Physics, POSTECH, Pohang 790-784, Korea,\\
and Department of Physics, Princeton University, Princeton, NJ 08540, USA 
}
 
\begin{abstract}
A local growth algorithm for a decagonal quasicrystal is presented. We
show that a perfect Penrose tiling (PPT) layer can be grown on a
decapod tiling layer by a three dimensional (3D) local rule
growth. Once a PPT layer begins to form on the upper layer, successive
2D PPT layers can be added on top resulting in a perfect decagonal
quasicrystalline structure in bulk with a point defect only on the
bottom surface layer. Our growth rule shows that an ideal quasicrystal
structure can be constructed by a local growth algorithm in 3D,
contrary to the necessity of non-local information for a 2D PPT
growth. 
\end{abstract}

%\\ 

\maketitle

%\section{Introduction} 

The announcement of icosahedral phase of alloys in
1984~\cite{Shechtman84} posed many puzzles. The first question was
what kind of arrangement of atoms could produce Bragg peaks with a
rotational symmetry forbidden to crystals. The quasiperiodic
translational order was proposed immediately 
as a candidate, and such materials began to be called 
quasiperiodic crystals or quasicrystals for short~\cite{Levine84}. 
However, the appearance of quasicrystals brings new puzzles: 
why and how the atoms can arrange themselves to 
have such order, and especially, how quasicrystals can 
grow with perfect quasiperiodic order has been a dilemma
since it seemingly requires non-local information while 
atomic interactions in metallic alloys are generally considered 
to be short ranged. 

There are currently two alternative pictures to describe
quasicrystals: energy-driven perfect quasiperiodic quasicrystals 
and entropy-driven random-tiling quasicrystals.
Accordingly, two alternative scenarios for the growth~\cite{Onoda88} 
of quasicrystals exist: matching-rule based, energy-driven 
growth, and finite-temperature entropy-driven growth~\cite{Grimm02I}.
A major criticism for the former approach has been that no 
local growth rules can produce a perfect quasicrystalline 
structure in 2D~\cite{Penrose89I,Dworkin95}.  Here, we show how to
overcome this obstacle in a 3D quasicrystals.  

\begin{figure}[t!] 
\includegraphics[width=7cm]{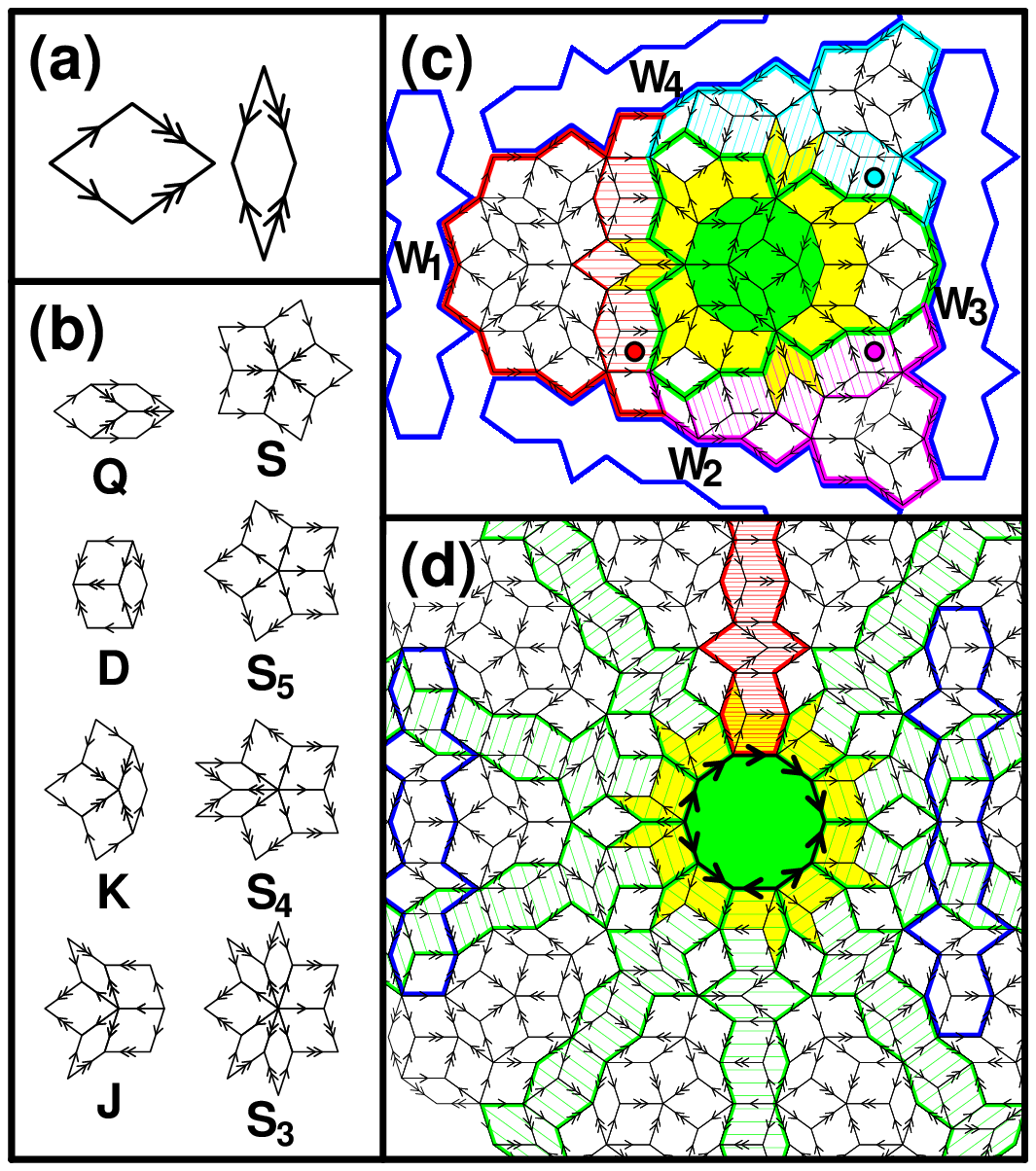} 
\caption[0]{(Color online) (a) Fat and skinny Penrose tiles with
arrows. (b) The eight ways of surrounding a vertex in a PPT. (c) Dead
surfaces encountered when a tiling is grown by Rule-L
from a
cartwheel decagon. See text for details. (d) A decapod tiling. 
Ten semi-infinite worms meet at the decapod decagon at the
center. 
From the yellow seed tiles, an infinite tiling (decapod tiling)
can be grown by Rule-L.
}
\label{f.osds}
\end{figure}

Penrose tiling~\cite{Penrose74} has been a basic template
for describing formation and structure of ideal quasicrystals. It can be
constructed from fat and thin rhombi with arrowed edges 
shown in Fig.~\ref{f.osds}(a). The infinite tiling consistent with 
arrow-matching rules is the Penrose tiling but they do not 
guarantee the growth of a perfect Penrose tiling (PPT) from 
a finite seed. Successive ``legal'' (obeying the arrow-matching
rules) additions of tiles to the surface of the 
already existing legal patch of tiles can produce defects. 
They usually occur after only a handful of tiles are added, 
and hence the arrow-matching rules cannot 
explain the long-range quasicrystalline order engendered by growth
kinetics.

There has been a great amount of discussion and a 
number of debates on the possibility of local growth 
algorithm for a PPT~\cite{Onoda88,Jaric89,Penrose89I,Dworkin95,
Ophuysen95,Socolar99I,Grimm02I}. The debates partially 
emerge from a different assumption on the growth processes 
at the surface, uniform growth and preferential 
growth. In the former, the growth occurs at any surface site 
with the same attaching probability, while it occurs with 
different attaching probabilities in the latter.  In 1988, 
Penrose proved that a PPT cannot be grown by local rules 
with uniform growth by showing that ``deceptions'' are 
unavoidable~\cite{Penrose89I} where a deception is a legal patch
which cannot be found in a PPT~\cite{Penrose89I,Dworkin95}.  In the
same year, Onoda \etal introduced a preferential growth algorithm
which can avoid deception by local rules called ``vertex rules''~\cite{Onoda88}. 
However, vertex rule growth stops at a ``dead surface'' and 
non-local information or arbitrarily small growth rates are 
required to be an infinite PPT. Yet, their growth algorithm 
is believed to provide methods to grow the most ideal 
quasicrystalline structures with local information. If an 
initial seed contains a special kind of defect, called
``decapod''~\cite{Gardner77}, 
Onoda \etal showed that the seed can be
grown to an almost PPT (whose only defect is the initial 
decapod defect)~\cite{Onoda88}. A point defect in a 2D tiling growth 
usually implies a line defect in a 3D decagonal tiling 
growth. If we apply a solid-on-solid type growth~\cite{Jeong99SSR} so 
that a layer copies configuration of the one below, we get 
decagonal tiling consisting of identical layers with a decapod 
defect at the center of each layer. This line defect has 
been considered to be a minimum imperfection for the 3D 
decagonal quasicrystal structure from the local growth 
algorithm.

%%%%%%%%%%%%%%%%%%%%%%%%(Fig 1: OSDS)%%%%%%%%%%%%%%%%%%%%%%
%%%%%%%%%%%%%%%%%%%%%%%%%%%%%%%%%%%%%%%%%%%%%%%%%%%%%%%%%%%%

In this letter, we consider the growth of decagonal 
quasicrystals and present a local growth algorithm for 3D 
decagonal tiling which consists of PPT layers except the 
bottom layer. We use the two well known results of the 
Onoda \etal study on planar decagonal quasicrystal growth~\cite{Onoda88}. 
1. A local growth method around a ``cartwheel decapod" 
leads to dead surfaces, from which further growth 
of a PPT requires non-local information.  2. Infinite local 
growth is possible if it starts from an ``active decapod 
defect" but the resultant tiling contains the defect and is 
not a PPT. By combining an active decapod defect in the 
bottom layer with a cartwheel decapod in the second layer
and using the information of the underneath configuration, 
we make the growth continue beyond dead surfaces in 
the second, and subsequent layers. The bottom surface layer 
has a point defect but we can consider the overall structure 
as that of a perfect decagonal quasicrystal since deviations 
from the bulk layer structure are natural for the surface 
layer even for ideal crystal materials. 

Let us first discuss the growth rule in a 2D Penrose tiling.  
For the arrow-matching rule growth, a deception can be made 
as few as three tiles~\cite{Penrose89I}.  Since the growth process 
does not allow tiles to be removed, a deception (which is 
not a part of a PPT) cannot grow to a PPT, and we need a 
growth rule which allows no deceptions of any size for a 
PPT growth. We can avoid three tile deceptions by introducing 
a more restricted growth rule which allows only 
correct (subset of a PPT) three tile patches. However, the 
new growth rule can make a deception in a larger scale, for 
example, a three tile deception of inflated~\cite{Gardner77} tiles.  Since a
deception can be made in all scales of multiply inflated tile
sizes~\cite{Gardner77}, it is unavoidable for a local growth rule. The
absence of local rules for perfect tiling growth seems to be 
the case for general aperiodic tilings in 2D~\cite{Dworkin95}. Based on
this observation, Penrose even speculates that there may be 
a non-local quantum-mechanical ingredient to quasicrystal
growth~\cite{Penrose89I,Penrose02B}.

Onoda \etal proposed ``vertex rules'' which avoid an 
encounter of deceptions~\cite{Onoda88}. Here, a tile can be added only 
to a ``forced edge'' which admits only one way of adding a
tile for its end vertices to be consistent with any of the eight 
PPT vertex configurations shown in Fig.~\ref{f.osds}(b). In this 
Letter, their vertex rules will be called ``Rule-L'' and used 
for the ``lateral''-direction growth (for 3D decagonal tilings).
The problem of Rule-L is that the growth stops at a
finite size patch called a ``dead surface'' which consists of 
unforced edges.

There are special kinds of point-like defects, called ``decapod''
defects~\cite{Gardner77}, which can be an ideal seed to grow an 
almost PPT without encountering a dead surface~\cite{Onoda88}. A 
decapod is a decagon with single arrowed edges. Since 
there are 10 arrows, each of which can take two independent 
orientations, there are $2^{10}$ combinations of states. 
After eliminating rotations and reflections, we get 62 distinct 
decapods. We can tile inside the decagon legally for 
only one decapod, the cartwheel case and the rest of the 61
decapods are called decapod defects. One notable property 
of the decapods is that the outside of the decagon region 
can be legally tiled for all 62 cases. This can be easily 
understood from the fact that six semi-infinite worms and 
two infinite worms meet at the center cartwheel decagon
[shown by green tiles in ~\ref{f.osds}(c)] in a cartwheel tiling. If 
we remove the tiles in the center cartwheel decagon, two 
infinite worms become four semi-infinite worms, and we 
have ten semi-infinite worms which start at the perimeter 
of the center decagon. A decapod defect tiling is formed by 
flipping one or more of these ten semi-infinite worms. The 
arrows on the worm perimeter will still fit except a mismatch
at the decapod decagon perimeter. Figure~\ref{f.osds}(d) 
shows an example obtained by flipping the worm denoted 
by red hatched tiles. Among 61 decapod defects, there are 
51 ``active'' decapod defects which have at least three 
consecutive arrows of the same orientation on their
decagon perimeter~\cite{Dotera91I}. One can show that a patch
containing an active decapod defect is never enclosed by a dead 
surface~\cite{Ingersent90}.  

\begin{figure}[t!] 
\includegraphics[width=7cm]{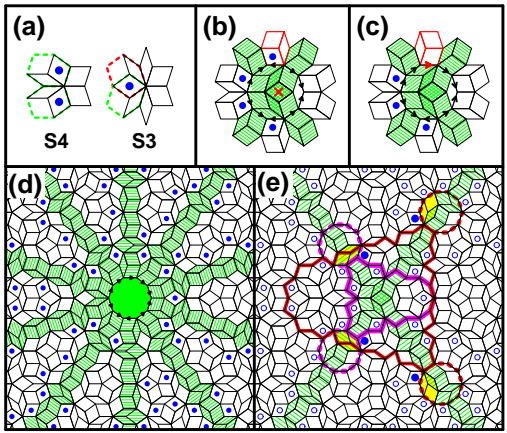} 
\caption[0]{(Color online) 
(a) Sticky sites, on which a tile can be attached vertically, are
indicated with blue circles. 
(b) Upper layer configuration of a (two layer) seed. It contains a
cartwheel decagon and ten hexagons attached to the decagon. 
The tiles denoted by blue circles are sticky site and the tile denoted
by X is the nucleate site.
(c) Lower layer configuration of the (two layer) seed. It contains an
active decapod decagon with five consecutive arrows of the same
orientation.  
(d) Sticky sites on a decapod tiling. All sticky sites are outside of
the ten semi-infinite worms. 
(e) Dead surfaces which contain the center cartwheel decagon 
in a cartwheel tiling. The two crossing infinite worms
of the cartwheel tiling always pass the two 72 degree corners of dead
surfaces. 
}
\label{f.vertical}
\end{figure}

Our 3D growth rules are constructed by observing that a 
cartwheel PPT and a decapod tiling can be different only in 
ten semi-infinite worms. Consider two layer growth from a 
(two layer) seed that contains a cartwheel decagon [yellow 
tiles in Fig.~\ref{f.osds}(c)] and a decapod defect [yellow tiles in 
Fig.~\ref{f.osds}(d)] at the upper and the lower layers respectively.
If each layer grows with Rule-L independently, the growth of 
the upper layer would stop at the red-purple-blue dead 
surface while the lower layers grow indefinitely. Now, we 
introduce a vertical growth rule so that a tile can be added 
at the dead surface of the upper layer properly. Note that 
the basic tiles for 3D decagonal tiling are rhombus prisms 
which have top and bottom faces as well as side faces.  By
vertical growth, we mean attaching tiles on the surface 
layer such that the bottom faces of the attached tiles contact
with the top faces of the surface layer tiles, while lateral growth
means attaching tiles to side faces at the perimeters.
We propose a vertical growth rule, ``Rule-V'' with which the lateral
growth rule, Rule-L produces a PPT on a decapod 
tiling. If a tile in a flipped worm of the decapod tiling (the 
lower layer) is copied by a vertical growth, a defect on the 
upper layer is inevitable. Our Rule-V is designed to avoid 
such a case and allow to attach tiles vertically only on the 
``sticky'' top faces, blue-circled fat tiles in $S_3$ and $S_4$ 
configurations shown in Fig.~\ref{f.vertical}(a)~\cite{loc.sticky}. 
They always form D-hexagons indicated by dotted lines. Such D-hexagons 
can lie only at the {\em end} of worms since the other (uncircled) 
tiles in $S_3$ or $S_4$ configurations prevent formation hexagons 
next to the circled tiles. Therefore, the sticky sites 
can be located only outside or at the ends of the semi-infinite 
worms as illustrated in Figs.~\ref{f.vertical}(b)-(e). One can 
further show that sticky sites are strictly outside of the 
semi-infinite worm if it is flipped since the flipping makes 
the vertices at the end be illegal (and therefore they cannot 
be $S_3$ or $S_4$). Therefore, no sticky sites are in the flipped 
worm, and hence Rule-V does not introduce a defect or 
deception for the layer that grows on a decapod tiling.  

%%%%%%%%%%%%%%%%%%%%%%%%(Fig 2: VERTICAL)%%%%%%%%%%%%%%%%%%%%%%
%%%%%%%%%%%%%%%%%%%%%%%%%%%%%%%%%%%%%%%%%%%%%%%%%%%%%%%%%%%%

Now we show that Rule-V is enough for the upper layer to 
grow beyond the dead surfaces when the growth starts from 
a proper seed. Our seed consists of two layer finite patches 
which include a cartwheel decagon and an active decapod 
decagon at the upper and the lower layers, respectively.  
Figures~\ref{f.vertical}(b) and~(c) show an example. The upper layer
seed [Fig.~\ref{f.vertical}(b)] consists of a cartwheel decagon and 10
D-hexagons. It covers all ends of the semi-infinite worms in 
the lower decapod seed [Fig.~\ref{f.vertical}(c)] which consists of a 
decapod decagon and 10 D-hexagons~\cite{loc.cover}. Let us first 
consider the properties of dead surfaces which contain 
the upper layer seed. By applying inflations to a cartwheel 
tiling, one can show that the dead surfaces, which contain 
the center cartwheel decagon, have two 72 degree corners and 
each corner is passed by an infinite worm [green worms in 
Fig.~\ref{f.vertical}(e)] of the cartwheel tiling~\cite{Socolar99I,Gardner77}. 
The D-hexagon at the 72-degree corner forces the next two hexagons
just outside of the dead surface (in the infinite worm direction) to
be D and Q. These two hexagons force a cartwheel decagon to 
form just outside of the corner as illustrated by (red and 
purple) dashed lines in Fig.~\ref{f.vertical}(e). A Q-decagon (denoted
by yellow tiles) in the dashed cartwheel decagon forces a tile 
just outside of the 72 degree corner [denoted by the solid green 
circles in Fig.~\ref{f.vertical}(e)] to be sticky. We call these
sticky sites as ``launching'' sites. The exact position of a launching
site depends on the orientation of the corner~\cite{Socolar99I} but
the patch can grow by Rule-L for both cases. The position of the 
launching site determines the orientation of the worm 
along the side lines of the dead surface making the edges 
at the dead surface become forced. Hence, the upper layer 
would grow to infinite by Rule-L if Rule-V guarantees tiles at 
the launching sites. This is the case when it grows on an 
active decapod tiling obtained by flipping a semi-infinite 
worm~\cite{loc.active} of a cartwheel tiling as shown in Fig.~\ref{f.vertical}(d).
Since neither crossing infinite-worms are flipped [compare 
Figs.~\ref{f.vertical}(d) and~(e)], the underneath tiles of the
launching sites will be always the sticky sites of the decapod tiling 
and tiles at the launching sites are guaranteed by Rule-V on 
the decapod tiling.   

For the completeness of the 3D decagonal quasicrystal 
growth, we need to provide the rule for the nucleation of an 
island (seed) from the third layer.  The physical process of 
the nucleation of an island on a PPT would be similar to 
that of a perfect crystal surface. High quality quasicrystals 
are grown when they grow slowly, or in other words, when 
the chemical potential of bulk quasicrystal is slightly less 
than that of the fluid phase. Therefore, adatoms or ``adtiles'' on 
a terrace would be unstable and probably diffuse 
on the terrace until they evaporate (i.e., go back to the fluid 
phase) or attach to preferential sites (forced or sticky
sites)~\cite{Jeong99SSR}. We believe that the chemical potentials
of the forced sites are less than those of sticky sites and adtiles
attach to forced sites for most cases. However, when the terrace 
forms a dead surface (or part of a dead surface), it is not 
easy for an adatom to find a forced site and it would attach 
on a sticky site, especially to a launching site whose 
chemical potential is expected to be lower than that of an 
isolated sticky site. Note that both forced and launching 
sites are at the perimeters of terraces and become irrelevant 
to adtiles on the middle of the terraces as they grow 
sufficiently large. It is then conceivable that two or more 
adtiles meet on a terrace and begin to form a new patch of 
the next layer before they arrive at the perimeter. With this 
physical process in mind, we allow a nucleation process in
our growth algorithm. The nucleation of an island can 
happen in cooperation of a quite large cluster of tiles. We 
choose the ``cartwheel seed'', a cartwheel decagon and the 
10 D-hexagons arranged as Fig.~\ref{f.vertical}(b), as such a cluster 
and introduce a nucleation site on it. The site X in the figure is 
called a ``nucleate site'' if its lateral neighboring tiles form 
a cartwheel seed and if it has a underneath tile~\cite{loc.phynucl}.
When a nucleate site is selected, we create a cartwheel seed on it. 

Let us summarize our growth mechanism for decagonal quasicrystals. 
It consists of three processes:
lateral growth by Rule-L, vertical growth by Rule-V, and the 
island nucleation (seed formation) for the new layer. 
Algorithmically, it is realized by the following steps:
1. Start with a two layer seed whose upper and lower 
layers contain a cartwheel decagon and an active decapod 
decagon, respectively. 2. Randomly choose a surface site. 
Check if it is a sticky, nucleate, or unsticky site when it is a 
top face. For a side face, check if it is a forced or unforced 
site. 3. Perform the vertical growth, nucleation, or lateral growth 
if the chosen site is a sticky, nucleate, or forced site,
respectively. Do nothing for unsticky (top face) or 
unforced (side face) site.  

For simplicity, we have chosen the unit attaching probability 
for all sticky, forced and nucleate sites. In real 
material, they probably have different attaching probabilities 
due to difference in their chemical potentials and 
attaching kinetics. We think that the attaching probabilities 
are different even among the forced sites (and among the 
sticky sites) since they depend on the local configurations. 
However, the nucleation probability would be much smaller 
than that of the attaching probability of a tile in 
any case since the former demands a cooperation of many 
tiles. Slow process of nucleation implies a layer by 
layer growth for a perfect decagonal quasicrystal~\cite{loc.nucleation}.
It is beyond the scope of this Letter to predict the growth 
kinetics of real quasicrystals since it requires knowing
atomic cluster structures corresponding to each type of 
tiles as well as the kinetic parameters of atomic attachment 
of real materials. 

Our growth algorithm has a couple of limitations. First, 
it can produce only one kind of PPT, a cartwheel tiling. 
Second, the seed must include a decapod defect. However, 
a decapod defect may form under quite general conditions. 
It is believed that every possible hole surrounded by an
arrow-matched Penrose tiling is equivalent to a decapod
hole~\cite{Gardner77}. The bottom layer, which may grow under 
structurally different environment, is natural place to have 
such defect. Our algorithm shows that PPT is possible from 
the second layer if the defect can be surrounded by legal 
tiles. Once a PPT layer begins to form on the second layer,
our growth algorithm produces PPT layers easily from the 
third layer.  We hope that the present work stimulates 
studies on 3D growth rules for real quasicrystals.

%\begin{acknowledgments}
We would like thank P. J. Steinhardt and M. Rechtsman for 
valuable comments. This work was supported by the Korea Research
Foundation Grant (KRF-2005-015-C00169).
%\end{acknowledgments}

%\bibliography{All,loc}

\begin{thebibliography}{10}
\bibitem{loc.sticky}
We introduce the concept of sticky sites as a mathematical devise to avoid the
  mistakes of copying a tile in a flipped worm but it may mimic the real growth
  process of quasicrystals. Recently, Fourn\'ee \etal observed ``traps'' or
  sticky sites on which adatoms are easily captured for some quasicrystal
  surfaces~\cite{Fournee03}.
\bibitem{loc.cover}
Covering of all ends of the semi-infinite worms further ensures that all sticky
  sites in the decapod tiling stay outside of the semi-infinite worms.
\bibitem{loc.active}
From the Fig~\ref{f.vertical}(b), one can see that an active decapod tiling is
  obtained by flipping a semi-infinite worm in a cartwheel tiling while
  flipping a half of an infinite worm results in an inactive decapod tiling.
\bibitem{loc.phynucl}
The requirement of the underneath tiles is added to ensure that the nucleation
  happens from the third layer. In real growth, this requirement may not be
  needed. Nucleation is likely to happen only on the top layer (hence from the
  third layer) which can be large enough to wait the slow necleation process.
\bibitem{loc.nucleation}
The sticky sites (Fig.~\ref{f.vertical}(a)) are formed only on a compact
  cluster of tiles. Therefore, the vertical growth by Rule-V, which produces
  isolated tiles, cannot be continue more than one layer height without
  nucleation process.
\end{thebibliography}
%\bibliographystyle{prsty}

\end{document}